\documentclass[twocolumn]{aastex631}

\usepackage{amsmath, mathrsfs, amssymb}
\usepackage{graphicx}
\usepackage{rotating}
\usepackage{morefloats}
\usepackage{color}
\usepackage{enumitem}
\usepackage{verbatim}
\usepackage{hyperref}
\usepackage{booktabs}
\usepackage{xfrac}
\usepackage{extarrows}

\renewcommand{\(}[0]{\left(}
\renewcommand{\)}[0]{\right)}
\renewcommand{\[}[0]{\left[}
\renewcommand{\]}[0]{\right]}

\newcommand{\deriv}[2]{\frac{\mathrm{d}#1}{\mathrm{d}#2}}

\renewcommand{\d}[1]{\mathrm{d}#1}

\setlist[itemize]{leftmargin=*}

\definecolor{darkblue}{RGB}{0, 0, 128}
\definecolor{gimppurple}{HTML}{AD26FB}

\begin{document}
\label{firstpage}

\shorttitle{Thermalization reverberation contemplation}
\shortauthors{Salvesen (2022)}

\title{\Large{An Electron-Scattering Time Delay in Black Hole Accretion Disks}}

\correspondingauthor{Greg Salvesen}
\email{gregsalvesen@gmail.com}

\author[0000-0002-9535-4914]{Greg Salvesen}
\affil{XCP-8, Los Alamos National Laboratory, P.O. Box 1663, Los Alamos, NM 87545, USA.}
\affil{Center for Theoretical Astrophysics, Los Alamos National Laboratory, Los Alamos, NM 87545, USA.}

\begin{abstract}
Universal to black hole X-ray binaries, the high-frequency soft lag gets longer during the hard-to-intermediate state transition, evolving from ${\lesssim}1~{\rm ms}$ to ${\sim}10~{\rm ms}$. The soft lag production mechanism is thermal disk reprocessing of non-thermal coronal irradiation. X-ray reverberation models account for the light-travel time delay external to the disk, but assume instantaneous reprocessing of the irradiation inside the electron scattering-dominated disk atmosphere. We model this neglected \textit{scattering time delay} as a random walk within an $\alpha$-disk atmosphere, with approximate opacities. To explain soft lag trends, we consider a limiting case of the scattering time delay that we dub the \textit{thermalization time delay}, $t_{\rm th}$; this is the time for irradiation to scatter its way down to the effective photosphere, where it gets thermalized, and then scatter its way back out. We demonstrate that $t_{\rm th}$ plausibly evolves from being inconsequential for low mass accretion rates $\dot{m}$ characteristic of the hard state, to rivaling or exceeding the light-travel time delay for $\dot{m}$ characteristic of the intermediate state. However, our crude model confines $t_{\rm th}$ to a narrow annulus near peak accretion power dissipation, so cannot yet explain in detail the anomalously long-duration soft lags associated with larger disk radii. We call for time-dependent models with accurate opacities to assess the potential relevance of a scattering delay.
\end{abstract}


\section{Introduction}
\label{sec:intro}
Imaging the X-ray-emitting regions of Galactic black hole (BH) X-ray binaries (XRBs) requires nano-arcsecond angular resolution; even a multi-spacecraft X-ray interferometer comes up short by factors of 1,000 \citep{Uttley2021}. But analogous to how a sound echo reveals a distance, a light echo (or `reverberation lag') reveals a size scale in a system that we could never image. By leveraging precision X-ray timing, XRB reverberation lags can map out the X-ray-emitting components that channel gas onto BHs, roughly divided into a `soft' thermal accretion disk and a `hard' non-thermal corona. When the corona is very active, its irradiation provokes a delayed reaction in the disk observed as a `soft lag', meaning that variability patterns in the soft photons lag behind those in the hard photons. In the parlance of XRB reverberation, `high-frequency' ($\gtrsim 1~{\rm Hz}$) refers to variability on short time scales ($\lesssim 1~{\rm s}$), which display soft (${\lesssim}1~{\rm keV}$) lags (${\sim}1$--$10~{\rm ms}$) attributed to thermal reprocessing of coronal irradiation by the disk atmosphere.

\begin{figure*}[!htb]
    \centering
    \vspace{-0mm}
    \includegraphics[width=1.0\textwidth]{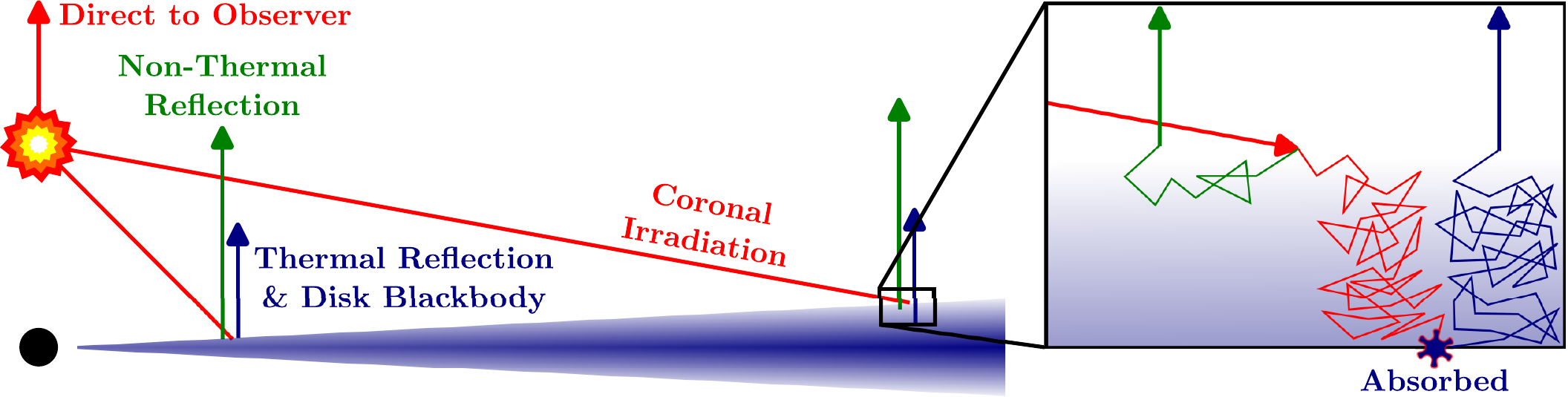}
    \vspace{-4mm}
    \caption{Reverberation lags are time delays between changes in the direct coronal emission and the corresponding variations in the disk-reprocessed emission. At the long-duration limit of our proposed scattering time delay is the thermalization time delay $t_{\rm th}$, which is the time taken by the scattering random walk during the thermalization process within the disk atmosphere.}
    \vspace{-0mm}
    \label{fig:lightecho}
\end{figure*}

Observed high-frequency soft lag durations are of order the light-travel time from the corona to the disk \citep{Uttley2011}, confirming the `thermal' reverberation lag proposed by \citet{WilkinsonUttley2009}. However, XRB reverberation lag models based solely on light-travel time delays systematically under-predict the high-frequency soft lag, which universally increases ${\sim}10$-fold during the hard-to-intermediate state transition \citep{Wang2022}.\footnote{Complicating matters, the soft lag initially gets shorter in the rising hard state, interpreted as coronal contraction \citep{Kara2019} or decreasing inner disk truncation \citep{DeMarco2021}.} Interpretations of these long-duration soft lags appeal to coronal expansion/ejection to increase the corona-to-disk light-travel time \citep{Wang2021, Wang2022}, under the assumption that the coronal irradiation gets reprocessed instantaneously within the disk atmosphere.
 
We hypothesize a \textit{scattering time delay} contribution to X-ray reverberation lags in BH XRBs, reducing the need for a contribution from coronal expansion and/or inner disk truncation. Figure \ref{fig:lightecho} outlines the basic idea, whereby the cumulative effect of individual Compton scatterings can rival the light-travel time delay, because electron scattering is the dominant opacity source in the disk atmosphere \citep[e.g.,][]{ShakuraSunyaev1973}.

A BH XRB spectrum consists of three components: direct disk blackbody emission, direct coronal continuum emission, and `reflection' produced from disk reprocessing of coronal irradiation. Reflection spectral models include physical processes that produce both soft/thermal and hard/non-thermal responses \citep[e.g.,][]{Garcia2010}, which X-ray reverberation studies assume to be simultaneous despite their different depths of formation in the disk atmosphere. Due to its association with the anomalously long-duration high-frequency soft lags, we focus on the thermal component of the reflection spectrum. In principle, the non-thermal component also experiences a scattering time delay, albeit shorter due to its shallower depth of formation.

In this Letter, we model the \textit{thermalization time delay} (\S\ref{sec:thermdelay}), which is an application of the more general electron-scattering time delay. We demonstrate the potential of this effect to explain the increasing duration of high-frequency soft lags from BH XRBs during their low-to-intermediate state transition (\S\ref{sec:results}). After discussing the role of a scattering time delay and its impact in a broader context (\S\ref{sec:disc}), we conclude (\S\ref{sec:sumconc}).

\section{Thermalization Time Delay Model}
\label{sec:thermdelay}
Let's derive an expression for the thermalization time delay $t_{\rm th}$ in terms of disk mid-plane quantities, which we can then evaluate with an $\alpha$-disk model.\footnote{Our derivation follows the spirit of \citet[][\S4.1]{BegelmanPringle2007} and \citet[][Appendix A and \S4.1]{SalvesenMiller2021}.} We approximate the thermalization time delay as
\begin{equation}
t_{\rm th} \sim \frac{2 N \ell}{c}, \label{eqn:t_th_1}
\end{equation}
by considering the journey of a typical irradiating photon that enters the electron scattering-dominated disk atmosphere. Traveling at the speed of light $c$, this photon experiences $N$ scattering events, with a mean free path $\ell$ between scatterings, before penetrating down to the effective photosphere to be absorbed and re-emitted with a lower energy. This thermalized photon then scatters its way back out of the atmosphere; hence, the factor of two in our estimate for $t_{\rm th}$ in equation \eqref{eqn:t_th_1}. We ignore photon energy-dependent differences on $N$ and $\ell$ between the ingoing and outgoing random walks.

For the densities and temperatures relevant to our problem, the electron scattering opacity is constant to good approximation throughout the disk atmosphere,\footnote{We assume a fully-ionized gas for $\kappa_{\rm es}$, but not for $\kappa_{\rm ff}$ and $\kappa_{\rm bf}$. This inconsistency is negligible for our application of a disk atmosphere with trace amounts of highly-ionized metals.}
\begin{equation}
\kappa_{\rm es} = \frac{\sigma_{\rm T}}{m_{\rm p}} \[X + \frac{1}{2} Y + \frac{1}{2} Z\], \label{eqn:kappa_es}
\end{equation}
with Thomson scattering cross section $\sigma_{\rm T}$ and proton mass $m_{\rm p}$. We assume solar abundances for the mass fractions of hydrogen ($X = 0.744$), helium ($Y = 0.242$), and metals ($Z = 0.014$), where $X + Y + Z = 1$ \citep{Asplund2021}. We assume a Kramers law for the free-free and bound-free opacities \citep[e.g.,][]{Schwarzschild1958},
\begin{align}
\kappa_{\rm ff} &\sim \[ 4 \times 10^{22}~{\rm cm^{2} / g}\] \( 1 + X \) \( 1 - Z \) \rho T^{-7/2} \label{eqn:kramers_ff} \\
\kappa_{\rm bf} &\sim \[ 4 \times 10^{25}~{\rm cm^{2} / g} \] \( 1 + X \) Z \rho T^{-7/2}, \label{eqn:kramers_bf}
\end{align}
which are functions of gas density $\rho$ [${\rm g/cm^{3}}$] and gas temperature $T$ [${\rm K}$]. The total absorption opacity (ignoring bound-bound) is $\kappa_{\rm th} = \kappa_{\rm ff} + \kappa_{\rm bf}$, or
\begin{equation}
\kappa_{\rm th} \sim \[ 4 \times 10^{25}~{\rm cm^{2} / g} \] \( 1 + X \) \( Z + 10^{-3} \) \rho T^{-7/2}. \label{eqn:kramers_th}
\end{equation} 

Random walk expressions for a combined scattering and absorbing medium \citep[e.g.,][]{RybickiLightman1979} relate the scattering quantities $N$ and $\ell$ to conditions at the effective photosphere (denoted by an asterisk),\footnote{Because $N \sim f_{\rm col}^{8}$, we expect a typical photon experiences tens to hundreds of scattering events during the thermalization process.}
\begin{align}
N &\sim \tau_{\ast}^{2} \label{eqn:N} \\
\ell &\sim \frac{\ell_{\ast}}{\sqrt{N}}. \label{eqn:ell}
\end{align}
A typical photon enters the disk atmosphere and traverses a vertical displacement $\ell_{\ast}$ down to the effective photosphere, where the optical depth is
\begin{equation}
\tau_{\ast} \equiv \int_{0}^{m_{\ast}} \( \kappa_{\rm es} + \kappa_{\rm th} \) \d{m} \gg 1. \label{eqn:tau_ast_1}
\end{equation}
Here, $m_{\ast}$ is the mass depth of the effective photosphere  and corresponds to where the effective ``optical depth'' $\tau_{\rm eff}^{\ast} \equiv \tau_{\rm eff}\( m_{\ast} \) = 1$, whose general expression is \citep[e.g.,][]{Davis2005}
\begin{equation}
\tau_{\rm eff}\( m \) \equiv \int_{0}^{m} \[ 3 \kappa_{\rm th} \( \kappa_{\rm es} + \kappa_{\rm th} \) \]^{1/2} \d{m^{\prime}}. \label{eqn:tau_eff}
\end{equation}
The mass depth relates to the vertical coordinate by $m\( z \) = \int_{z}^{\infty} \rho \d{z}$, where the mid-plane location is $z = 0$, or $m_{\rm d} = \Sigma_{\rm d} / 2$, with $\Sigma_{\rm d}$ the disk surface mass density.

Assuming electron scattering is the dominant opacity source, the effective photosphere optical depth becomes
\begin{equation}
\tau_{\ast} \simeq \tau_{\rm es}^{\ast} \sim \kappa_{\rm es} \rho_{\ast} \ell_{\ast}, \label{eqn:tau_ast_2}
\end{equation}
where $\rho_{\ast}$ is the gas density at the effective photosphere. We can express $\tau_{\rm es}^{\ast}$ in terms of the scattering and absorption opacities using the definition of the effective photosphere,
\begin{equation}
\tau_{\rm eff}^{\ast} \equiv 1 \sim \tau_{\rm es}^{\ast} \( \frac{\kappa_{\rm th}^{\ast}}{\kappa_{\rm es}} \)^{1/2}. \label{eqn:tau_eff_ast}
\end{equation}
This relation follows from combining the approximations $\tau_{\rm es}^{\ast} \sim \kappa_{\rm es} m_{\ast}$ and $\tau_{\rm eff}^{\ast} \equiv 1 \sim \( \kappa_{\rm th}^{\ast} \kappa_{\rm es} \)^{1/2} m_{\ast}$, which come from equations \eqref{eqn:tau_ast_1} and \eqref{eqn:tau_eff} assuming $\kappa_{\rm es} \gg \kappa_{\rm th}^{\ast}$.

Using equations \eqref{eqn:N}--\eqref{eqn:tau_eff_ast}, we express the $t_{\rm th}$ estimate from equation \eqref{eqn:t_th_1} in terms of the gas density and absorption opacity at the effective photosphere,
\begin{equation}
t_{\rm th} \sim \frac{2}{c} \frac{1}{\kappa_{\rm th}^{\ast} \rho_{\ast}}. \label{eqn:t_th_2}
\end{equation}

Calculating the ratio of the absorption opacity at the effective photosphere to that at the disk mid-plane,
\begin{equation}
\frac{\kappa_{\rm th}^{\ast}}{\kappa_{\rm th}^{\rm d}} = \( \frac{\rho_{\ast}}{\rho_{\rm d}} \) \( \frac{T_{\ast}}{T_{\rm d}} \)^{-7/2}, \label{eqn:kappa_th}
\end{equation}
and then substituting $\kappa_{\rm th}^{\ast}$ into equation \eqref{eqn:t_th_2} gives,
\begin{equation}
t_{\rm th} \sim \frac{2}{c} \frac{1}{\kappa_{\rm th}^{\rm d} \rho_{\rm d}} \( \frac{T_{\ast}}{T_{\rm d}} \)^{7/2} \( \frac{\rho_{\ast}}{\rho_{\rm d}} \)^{-2}, \label{eqn:t_th_3}
\end{equation}
where a subscript or superscript `${\rm d}$' denotes a quantity evaluated at the disk mid-plane.

We now spend a significant amount of time approximating $T_{\ast} / T_{\rm d}$, the ratio of the gas temperature at the effective photosphere to that at the disk mid-plane. The simplified result will be $T_{\ast} / T_{\rm d} \sim \(\tau_{\rm es}^{\ast} / \tau_{\rm es}^{\rm d}\)^{1/4}$, but we choose to consider coronal irradiative heating and disk surface layer dissipation to examine their effects on $t_{\rm th}$.

Assuming heat transport by radiative diffusion only, and assuming local thermodynamic equilibrium (LTE), the equation for the outgoing radiative energy flux is
\begin{equation}
F_{\rm r} = c \deriv{P_{\rm r}}{\tau} \xlongrightarrow{\text{LTE}} F_{\rm r} = \frac{4}{3} \deriv{\(\sigma_{\rm r} T^{4}\)}{\tau}, \label{eqn:raddiff}
\end{equation}
where $P_{\rm r}$ is the radiation pressure and $\sigma_{\rm r}$ is the Stefan-Boltzmann constant. Integrating equation \eqref{eqn:raddiff} starting from the effective photosphere $\(\tau = \tau_{\ast}\)$, let's go with the light, up to the highest height $\(\tau = 0\)$,
\begin{align}
\int_{\sigma_{\rm r} T_{\ast}^{4}}^{0} \d{\(\sigma_{\rm r} T^{4}\)} &= \frac{3}{4} \int_{\tau_{\ast}}^{0} F_{\rm r} \d{\tau} \nonumber \\
\sigma_{\rm r} T_{\ast}^{4} &\sim \frac{3}{4} F_{\rm r}^{\ast} \tau_{\ast}. \label{eqn:poppins_star}
\end{align}
Now starting from the disk mid-plane ($\tau = \tau_{\rm d}$), let's go with the light, and keep on soaring, up through the atmosphere, up where there's photosphere ($\tau = \tau_{\ast}$),
\begin{align}
\int_{\sigma_{\rm r} T_{\rm d}^{4}}^{\sigma_{\rm r} T_{\ast}^{4}} \d{\(\sigma_{\rm r} T^{4}\)} &= \frac{3}{4} \int_{\tau_{\rm d}}^{\tau_{\ast}} F_{\rm r} \d{\tau} \nonumber \\
\sigma_{\rm r} T_{\ast}^{4} - \sigma_{\rm r} T_{\rm d}^{4} &\sim \frac{3}{4} \( F_{\rm r}^{\ast} \tau_{\ast} - F_{\rm r}^{\rm d} \tau_{\rm d} \). \label{eqn:poppins_disk}
\end{align}
In evaluating the integrals, we assumed an outgoing radiative flux $F_{\rm r}^{\rm d}$ emanates from the disk mid-plane and remains constant until reaching the effective photosphere, where we allow $F_{\rm r}^{\rm d}$ to suddenly increase to $F_{\rm r}^{\ast}$ to approximately account for coronal irradiative heating and/or disk surface layer dissipation. Using the definitions of the effective temperatures associated with the disk mid-plane and the effective photosphere,
\begin{align}
\sigma_{\rm r} \(T_{\rm eff}^{\rm d}\)^{4} &\equiv F_{\rm r}^{\rm d} = \(1 - f\) F_{\rm acc} \label{eqn:Teff_d} \\
\sigma_{\rm r} \(T_{\rm eff}^{\ast}\)^{4} &\equiv F_{\rm r}^{\ast} = F_{\rm acc} + \(1 - \beta\) F_{\rm irr}, \label{eqn:Teff_ast}
\end{align}
we partition $F_{\rm r}^{\rm d}$ and $F_{\rm r}^{\ast}$ into flux contributions from disk accretion $F_{\rm acc}$ and coronal irradiation $F_{\rm irr}$. Some fraction $f$ of the available accretion power gets dissipated in the disk surface layers \citep{SvenssonZdziarski1994}. Some fraction $\(1 - \beta\)$ of the irradiative flux gets absorbed in the disk surface layers, where $\beta$ is the disk surface albedo. We assume the dissipated accretion power and the absorbed coronal irradiation get reprocessed into thermal radiation at the effective photosphere.

Dividing equation \eqref{eqn:poppins_disk} by \eqref{eqn:poppins_star}, substituting equations \eqref{eqn:Teff_d} and \eqref{eqn:Teff_ast}, and recalling that electron scattering is the dominant opacity source, gives
\begin{equation}
\frac{T_{\ast}}{T_{\rm d}} \sim \(\frac{F_{\rm r}^{\ast}}{F_{\rm r}^{\rm d}} \frac{\tau_{\ast}}{\tau_{\rm d}}\)^{1/4} = \( \frac{\tau_{\rm es}^{\ast}}{\tau_{\rm es}^{\rm d}} \)^{1/4} \[\frac{1 + \(1 - \beta\) F_{\rm irr} / F_{\rm acc}}{1 - f}\]^{1/4}, \label{eqn:T_ratio}
\end{equation}
which reduces to $T_{\ast} / T_{\rm d} \sim \(\tau_{\rm es}^{\ast} / \tau_{\rm es}^{\rm d}\)^{1/4}$ in the absence of both absorbed coronal irradiation ($\beta = 1$) and accretion power dissipation in the disk surface layers ($f = 0$).

Using equations \eqref{eqn:tau_eff_ast} and \eqref{eqn:kappa_th}, equation \eqref{eqn:T_ratio} becomes
\begin{align}
\frac{T_{\ast}}{T_{\rm d}} &\sim \(\tau_{\rm es}^{{\rm d}}\)^{-4/9} \( \frac{\kappa_{\rm es}}{\kappa_{\rm th}^{\rm d}} \)^{2/9} \( \frac{\rho_{\ast}}{\rho_{\rm d}} \)^{-2/9} \nonumber \\
&\times \[\frac{1 + \(1 - \beta\) F_{\rm irr} / F_{\rm acc}}{1 - f}\]^{4/9},
\end{align}
which we plug into equation \eqref{eqn:t_th_3} to obtain the desired expression for the thermalization time delay,
\begin{align}
t_{\rm th} &\sim \frac{2}{c} \kappa_{\rm es}^{7/9} \(\kappa_{\rm th}^{\rm d}\)^{-16/9} \(\tau_{\rm es}^{\rm d}\)^{-14/9} \rho_{\rm d}^{-1} \( \frac{\rho_{\ast}}{\rho_{\rm d}} \)^{-25/9} \nonumber \\
&\times \[\frac{1 + \(1 - \beta\) F_{\rm irr} / F_{\rm acc}}{1 - f}\]^{14/9}, \label{eqn:t_th_4}
\end{align}
written in terms of the disk mid-plane quantities $\tau_{\rm es}^{\rm d}$, $\kappa_{\rm th}^{\rm d}$, $\rho_{\rm d}$, and the density at the effective photosphere $\rho_{\ast}$. Note that $\rho_{\ast} / \rho_{\rm d} < 1$ provides a lower limit to $t_{\rm th}$.

\subsection{Disk Radial Structure}
\label{sec:diskstruct}
Assuming an $\alpha$-disk model \citep{ShakuraSunyaev1973, NovikovThorne1973}, we collect approximations of the vertically-averaged equations for hydrostatic equilibrium, angular momentum conservation, disk energy balance, and radiative diffusion, respectively \citep{SvenssonZdziarski1994, RiffertHerold1995},
\begin{align}
P_{\rm d} &= \rho_{\rm d} \frac{G M}{R^{3}} H_{\rm d}^{2} \frac{1}{\zeta} \frac{\mathscr{C}}{\mathscr{B}} \label{eqn:hydrostateq} \\
P_{\rm d} &= \frac{\dot{M}}{4 \pi \alpha H_{\rm d}} \(\frac{G M}{R^{3}}\)^{1/2} \frac{\mathscr{D} \mathscr{B}^{1/2} \mathscr{C}^{1/2}}{\mathscr{A}^{2}} \label{eqn:angmomcons} \\
F_{\rm r}^{\rm d} &= \(1 - f\) \frac{3 G M \dot{M}}{8 \pi R^{3}} \frac{\mathscr{D}}{\mathscr{B}} \label{eqn:enrgbal} \\
F_{\rm r}^{\rm d} &= c \frac{P_{\rm r}^{\rm d}}{\tau_{\rm d}} \frac{2}{\xi}, \label{eqn:raddiffavg}
\end{align}
adopting the \citet{ShakuraSunyaev1973} conventions $\zeta = 1$ and $\xi = 1$. We want to solve for the pressure scale height $H_{\rm d}$, gas density $\rho_{\rm d}$, optical depth $\tau_{\rm d}$, pressure $P_{\rm d}$, and gas temperature $T_{\rm d}$; as functions of the effective viscosity parameter $\alpha$ and dimensionless versions of the BH mass $m \equiv M / M_{\odot}$, disk radius $r \equiv R / R_{\rm g}$, and mass accretion rate $\dot{m} \equiv \dot{M} c^{2} / L_{\rm Edd}$. These come from scaling their dimensional quantities to the solar mass $M_{\odot}$, gravitational radius $R_{\rm g} = G M / c^{2}$, and Eddington luminosity $L_{\rm Edd} = 4 \pi G M c / \kappa_{\rm es}$. The radiative efficiency factor
\begin{equation}
\eta = 1 - \(1 - \frac{2}{3 r_{\rm in}}\)^{1/2}
\end{equation}
maps $\dot{m}$ to the Eddington-scaled disk luminosity $l_{\rm d} \equiv L_{\rm disk} / L_{\rm Edd} = \eta \dot{m}$. The BH spin parameter $a_{\ast}$ determines $\eta$ by assuming the inner disk radius $r_{\rm in} \equiv R_{\rm in} / R_{\rm g}$ coincides with the innermost stable circular orbit \citep{Bardeen1972}. For the change of variables $x \equiv r^{1/2}$, the $\(x, a_{\ast}\)$-dependent relativistic corrections are those defined in \citet{RiffertHerold1995},
\begin{equation}
\mathscr{A} \equiv 1 - \frac{2}{x^{2}} + \frac{a_{\ast}^{2}}{x^{4}}, \indent \mathscr{B} \equiv 1 - \frac{3}{x^{2}} + \frac{2 a_{\ast}}{x^{3}}, \indent \mathscr{C} \equiv 1 - \frac{4 a_{\ast}}{x^{3}} + \frac{3 a_{\ast}^{2}}{x^{4}}, \nonumber
\end{equation}
and
\begin{align}
\mathscr{D} &\equiv \frac{1}{x} \left[ x - x_{0} - \frac{3}{2} a_{\ast} \ln\(\frac{x}{x_{0}}\) \right. \nonumber \\
&- \frac{3 \(x_{1} - a_{\ast}\)^{2}}{x_{1} \(x_{1} - x_{2}\) \(x_{1} - x_{3}\)} \ln\(\frac{x - x_{1}}{x_{0} - x_{1}}\) \nonumber \\
&- \frac{3 \(x_{2} - a_{\ast}\)^{2}}{x_{2} \(x_{2} - x_{1}\) \(x_{2} - x_{3}\)} \ln\(\frac{x - x_{2}}{x_{0} - x_{2}}\) \nonumber \\
&\left.- \frac{3 \(x_{3} - a_{\ast}\)^{2}}{x_{3} \(x_{3} - x_{1}\) \(x_{3} - x_{2}\)} \ln\(\frac{x - x_{3}}{x_{0} - x_{3}}\) \right],
\end{align}
with $x_{0}$, $x_{1}$, $x_{2}$, $x_{3}$ as defined in \citet{PageThorne1974}.

For the X-ray-emitting inner disk regions dominated by radiation pressure, $P_{\rm d} = P_{\rm r}^{\rm d} = 4 \sigma_{\rm r} T_{\rm d}^{4} / \(3 c\)$, and electron scattering opacity, $\tau_{\rm d} = \tau_{\rm es}^{\rm d} = \kappa_{\rm es} \rho_{\rm d} H_{\rm d}$, solving equations \eqref{eqn:hydrostateq}--\eqref{eqn:raddiffavg} gives the mid-plane structure,
\begin{align}
H_{\rm d} &= \frac{3}{4} \frac{G M_{\odot}}{c^{2}} m \dot{m} \mathscr{C}^{-1} \mathscr{D} \zeta \xi \(1 - f\) \label{eqn:H} \\
\rho_{\rm d} &= \frac{64}{27} \frac{1}{\kappa_{\rm es}} \frac{c^{2}}{G M_{\odot}} \alpha^{-1} m^{-1} r^{3/2} \dot{m}^{-2} \nonumber \\
&\times \mathscr{A}^{-2} \mathscr{B}^{3/2} \mathscr{C}^{5/2} \mathscr{D}^{-2} \zeta^{-2} \[\xi \(1 - f\)\]^{-3} \label{eqn:rho} \\
\tau_{\rm es}^{\rm d} &= \frac{16}{9} \alpha^{-1} r^{3/2} \dot{m}^{-1} \nonumber \\
&\times \mathscr{A}^{-2} \mathscr{B}^{3/2} \mathscr{C}^{3/2} \mathscr{D}^{-1} \zeta^{-1} \[\xi \(1 - f\)\]^{-2} \label{eqn:tau_es} \\
P_{\rm d} &= \frac{4}{3} \frac{c^{2}}{\kappa_{\rm es}} \frac{c^{2}}{G M_{\odot}} \alpha^{-1} m^{-1} r^{-3/2} \nonumber \\
&\times \mathscr{A}^{-2} \mathscr{B}^{1/2} \mathscr{C}^{3/2} \[\zeta \xi \(1 - f\)\]^{-1} \label{eqn:P} \\
T_{\rm d} &= \( \frac{c^{3}}{\sigma_{\rm r} \kappa_{\rm es}} \frac{c^{2}}{G M_{\odot}} \)^{1/4} \alpha^{-1/4} m^{-1/4} r^{-3/8} \nonumber \\
&\times \mathscr{A}^{-1/2} \mathscr{B}^{1/8} \mathscr{C}^{3/8} \[\zeta \xi \(1 - f\)\]^{-1/4}. \label{eqn:T}
\end{align}

\subsection{Thermalization Time Delay}
\label{sec:t_th}
Our analytic form of the thermalization time delay,
\begin{align}
t_{\rm th} &\sim \[ 3.9 \times 10^{6}~{\rm s} \] \alpha^{25/9} m^{11/9} r^{-53/6} \dot{m}^{64/9} \( \frac{\rho_{\ast}}{\rho_{\rm d}} \)^{-25/9} \nonumber \\
&\times \mathscr{A}^{50/9} \mathscr{B}^{-103/18} \mathscr{C}^{-125/18} \mathscr{D}^{64/9} \zeta^{50/9} \[\xi \(1 - f\)\]^{89/9} \nonumber \\
&\times \[\frac{1 + \(1 - \beta\) F_{\rm irr} / F_{\rm acc}}{1 - f}\]^{14/9}, \label{eqn:thermdelay}
\end{align}
follows from inserting $\kappa_{\rm es}$, $\kappa_{\rm th}^{\rm d}$ (equations \ref{eqn:kappa_es}, \ref{eqn:kramers_th}) and $\rho_{\rm d}$, $\tau_{\rm es}^{\rm d}$, $T_{\rm d}$ (equations \ref{eqn:rho}, \ref{eqn:tau_es}, \ref{eqn:T}) into equation \eqref{eqn:t_th_4}. In this process, one discovers the radial disk structure scalings $t_{\rm th} \propto \(\tau_{\rm es}^{\rm d}\)^{-14/9} \rho_{\rm d}^{-25/9} T_{\rm d}^{56/9} \(\rho_{\ast} / \rho_{\rm d}\)^{-25/9}$ responsible for the extreme parameter sensitivities of our model.

To obtain non-relativistic versions of the preceding equations \citep[e.g.,][]{SvenssonZdziarski1994}, replace $\mathscr{A}$, $\mathscr{B}$, $\mathscr{C}$ with unity and $\mathscr{D}$ with $J\(r\) \equiv 1 - \(r_{\rm in} / r\)^{1/2}$, the familiar form used by \citet{ShakuraSunyaev1973}.

\subsection{Light-Travel Time Delay}
\label{sec:t_lt}
The light-travel time delay
\begin{equation}
t_{\rm lt}\(r, \phi\) = t_{\rm cd}\(r\) + t_{\rm do}\(r, \phi\) - t_{\rm co}\(r\),
\end{equation}
depends on disk radius $r \equiv R / R_{\rm g}$ and azimuthal angle\footnote{The point on a disk ring closest to the observer defines $\phi = 0$.} $\phi$, where the corona-to-disk, disk-to-observer, and corona-to-observer light-travel times are
\begin{align}
t_{\rm cd}\(R\) &= \frac{1}{c} \(R^{2} + H_{\rm c}^{2}\)^{1/2} \\
t_{\rm do}\(R, \phi\) &= \frac{1}{c} \[D^{2} + R^{2} - 2 D R \sin\(i\) \cos\(\phi\)\]^{1/2} \\
t_{\rm co} &= \frac{1}{c} \[D^{2} + H_{\rm c}^{2} - 2 D H_{\rm c} \cos\(i\)\]^{1/2}.
\end{align}
Here and throughout, we ignore light bending, adopt a `lamppost' corona at dimensionless height $h_{\rm c} \equiv H_{\rm c} / R_{\rm g}$ \citep{Matt1992}, and assume a razor-thin disk at inclination angle $i$ lying in the equatorial plane of the BH, whose distance from the observer is $D$. Because $R \ll D$ and $H_{\rm c} \ll D$, we can Taylor expand $t_{\rm do}\(r, \phi\)$ and $t_{\rm co}$ to get \citep[e.g.,][equation B4]{Mastroserio2018}
\begin{equation}
t_{\rm lt}\(r, \phi\) \simeq \frac{R_{\rm g}}{c} \[ \(r^{2} + h_{\rm c}^{2}\)^{1/2} - r \sin\(i\) \cos\(\phi\) + h_{\rm c} \cos\(i\) \]. \label{eqn:t_lt}
\end{equation}

\subsection{Coronal Irradiation and Disk Accretion Fluxes}
\label{sec:Firr_Facc}
Considering a time-steady dimensionless coronal luminosity $l_{\rm c} \equiv L_{\rm c} / L_{\rm Edd}$, the radial profile of coronal flux irradiating the disk is
\begin{equation}
F_{\rm irr}\(r\) = \frac{L_{\rm c}}{4 \pi R_{\rm g}^{2}} \frac{h_{\rm c}}{\(h_{\rm c}^{2} + r^{2}\)^{3/2}} \frac{2 \pi r R_{\rm g}^{2}}{\d{A_{\rm ring}} / \d{r}}, \label{eqn:Firr}
\end{equation}
which follows from integrating both sides of \citet[][equation 14]{Ingram2019}, ignoring light bending and redshifting. As measured in the disk frame, an annulus of width $\d{r}$ has area \citep[e.g.,][equation 9]{Bardeen1972, WilkinsFabian2012}
\begin{equation}
\frac{\d{A_{\rm ring}}}{\d{r}} = \frac{r^{1/4} \(r^{3/2} \pm a_{\ast}\)}{\(r^{3/2} - 3 r^{1/2} \pm 2 a_{\ast}\)^{1/2}} 2 \pi R_{\rm g}^{2}, \label{eqn:dAring_dr}
\end{equation}
where the upper (lower) signs refer to prograde (retrograde) orbits. The entirety of the irradiating flux gets reprocessed by the disk. However, to isolate the thermal response responsible for the soft lag, equation \eqref{eqn:Teff_ast} only considers the portion $\(1 - \beta\) F_{\rm irr}$ that gets thermalized after a time delay $t_{\rm th}$ \citep[e.g.,][]{Cackett2007}.

In addition to this thermally reprocessed flux, steady-state disk accretion liberates a gravitational energy flux\footnote{Our model assumes $F_{\rm acc}$ gets dissipated as a radiative energy flux; a fraction $f$ in the disk surface and $\(1 - f\)$ in the mid-plane.}
\begin{equation}
F_{\rm acc}\(r\) = \frac{3 G M \dot{M}}{8 \pi \(r R_{\rm g}\)^{3}} \frac{\mathscr{D}}{\mathscr{B}}. \label{eqn:Facc}
\end{equation}
 
Dividing equation \eqref{eqn:Firr} by \eqref{eqn:Facc}, the coronal irradiative flux relative to the disk accretion radiative flux is
\begin{equation}
\frac{F_{\rm irr}\(r\)}{F_{\rm acc}\(r\)} = \frac{2}{3} \frac{l_{\rm c}}{l_{\rm d} / \eta} \frac{\mathscr{B}}{\mathscr{D}} \frac{h_{\rm c} r^{3}}{\(r^{2} + h_{\rm c}^{2}\)^{3/2}} \frac{2 \pi r R_{\rm g}^{2}}{\d{A_{\rm ring}} / \d{r}}. \label{eqn:Firr_Facc}
\end{equation}

\subsection{Color Correction}
\label{sec:f_col}
The dominance of electron scattering opacity leads to an optical depth at the effective photosphere $\tau_{\ast} \gg 1$ (equation \ref{eqn:tau_ast_1}). Because $\tau_{\ast} \ne 1$, the actual temperature at the effective photosphere $T_{\ast}$ is \textit{not} equivalent to the corresponding effective temperature $T_{\rm eff}^{\ast}$. Their relationship $T_{\ast} = f_{\rm col} T_{\rm eff}^{\ast}$ through the color correction $f_{\rm col} \sim \(3 \tau_{\ast} / 4\)^{1/4}$ follows from comparing equations \eqref{eqn:poppins_star} and \eqref{eqn:Teff_ast}. To be consistent with our thermalization time delay model, we can follow the same methodology to express $f_{\rm col}$ as a function of $\alpha$-disk model parameters,
\begin{align}
f_{\rm col} &\sim \[ 2.9 \times 10^{1} \] \alpha^{2/9} m^{1/36} r^{-11/12} \dot{m}^{23/36} \( \frac{\rho_{\ast}}{\rho_{\rm d}} \)^{-2/9} \nonumber \\
&\times \mathscr{A}^{4/9} \mathscr{B}^{-19/36} \mathscr{C}^{-5/9} \mathscr{D}^{23/36} \zeta^{4/9} \[\xi \(1 - f\)\]^{31/36} \nonumber \\
&\times \[\frac{1 + \(1 - \beta\) F_{\rm irr} / F_{\rm acc}}{1 - f}\]^{7/36}, \label{eqn:f_col}
\end{align}
whose effect is to spectrally harden the disk continuum. Importantly, the color temperature is the observationally relevant quantity, \textit{not} the effective temperature.

\section{Results}
\label{sec:results}

\begin{figure}[!t]
    \centering
    \vspace{-0mm}
    \includegraphics[width=0.495\textwidth]{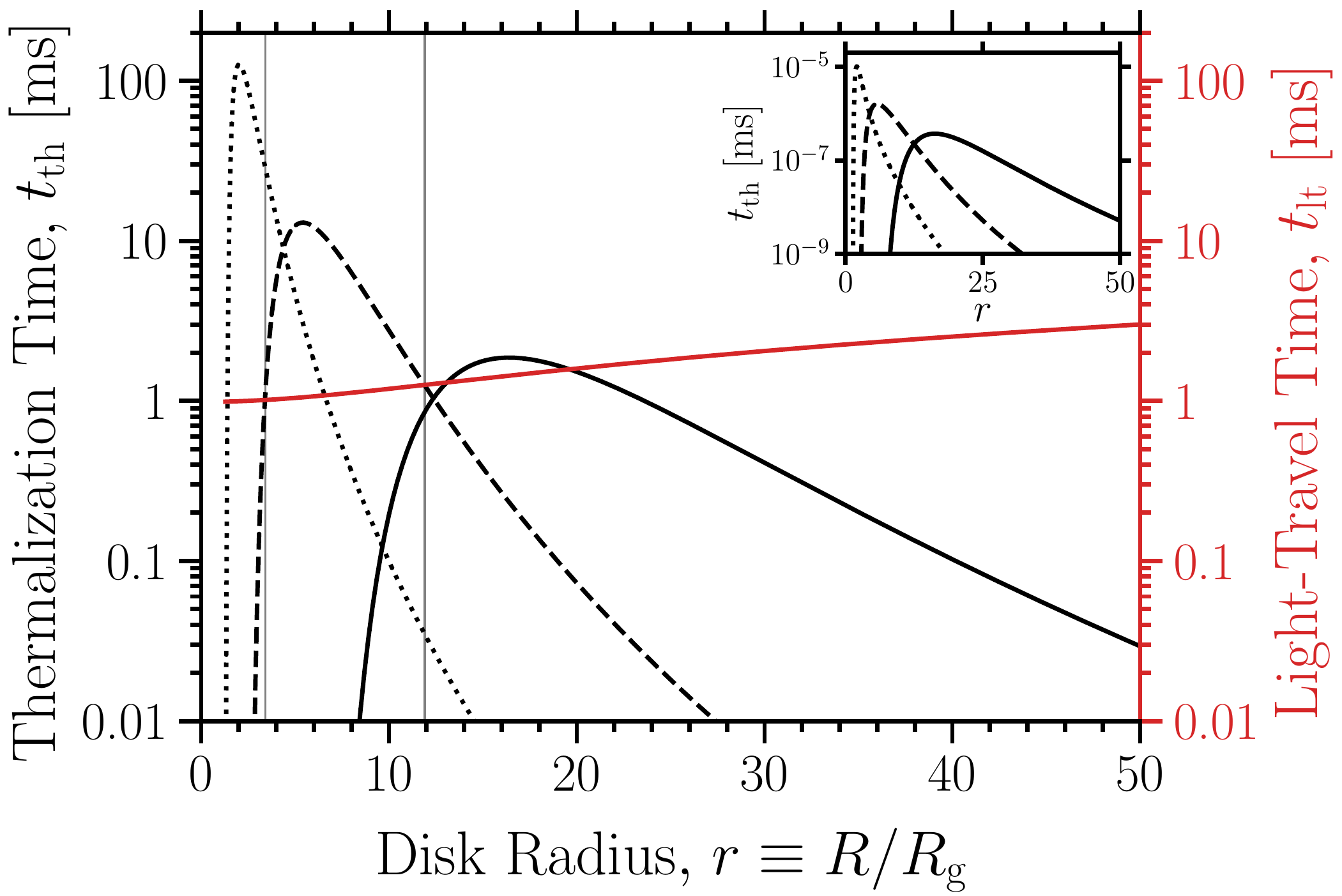}
    \vspace{-6mm}
    \caption{\textit{Left/black axis}: Thermalization time delay $t_{\rm th}$ in milliseconds, as a function of disk radius in gravitational units. Parameter choices are deliberately suggestive, but represent an XRB in the intermediate state ($l_{\rm d} = 0.2$) with a BH mass $m = 10$ and BH spin $a_{\ast} = 0$ (\textit{solid}), 0.9 (\textit{dashed}), 0.998 (\textit{dotted}). We set $\alpha = 0.2$, $\rho_{\ast} / \rho = 0.1$, $\beta = 0.5$, $l_{\rm c} = 0.05$, $h_{\rm c} = 10$. \textit{Right/red axis}: Light-travel time delay $t_{\rm lt}$ for a face-on observer ($i = 0^{\circ}$). Thin vertical lines mark the narrow annulus where $t_{\rm th} > t_{\rm lt}$ for the $a_{\ast} = 0.9$ case. \textit{Inset}: For a representative low/hard state ($l_{\rm d} = 0.02$), $t_{\rm th}$ is negligible.}
    \vspace{-0mm}
    \label{fig:therm}
\end{figure}

\begin{figure}[!t]
    \centering
    \vspace{-0mm}
    \includegraphics[width=0.495\textwidth]{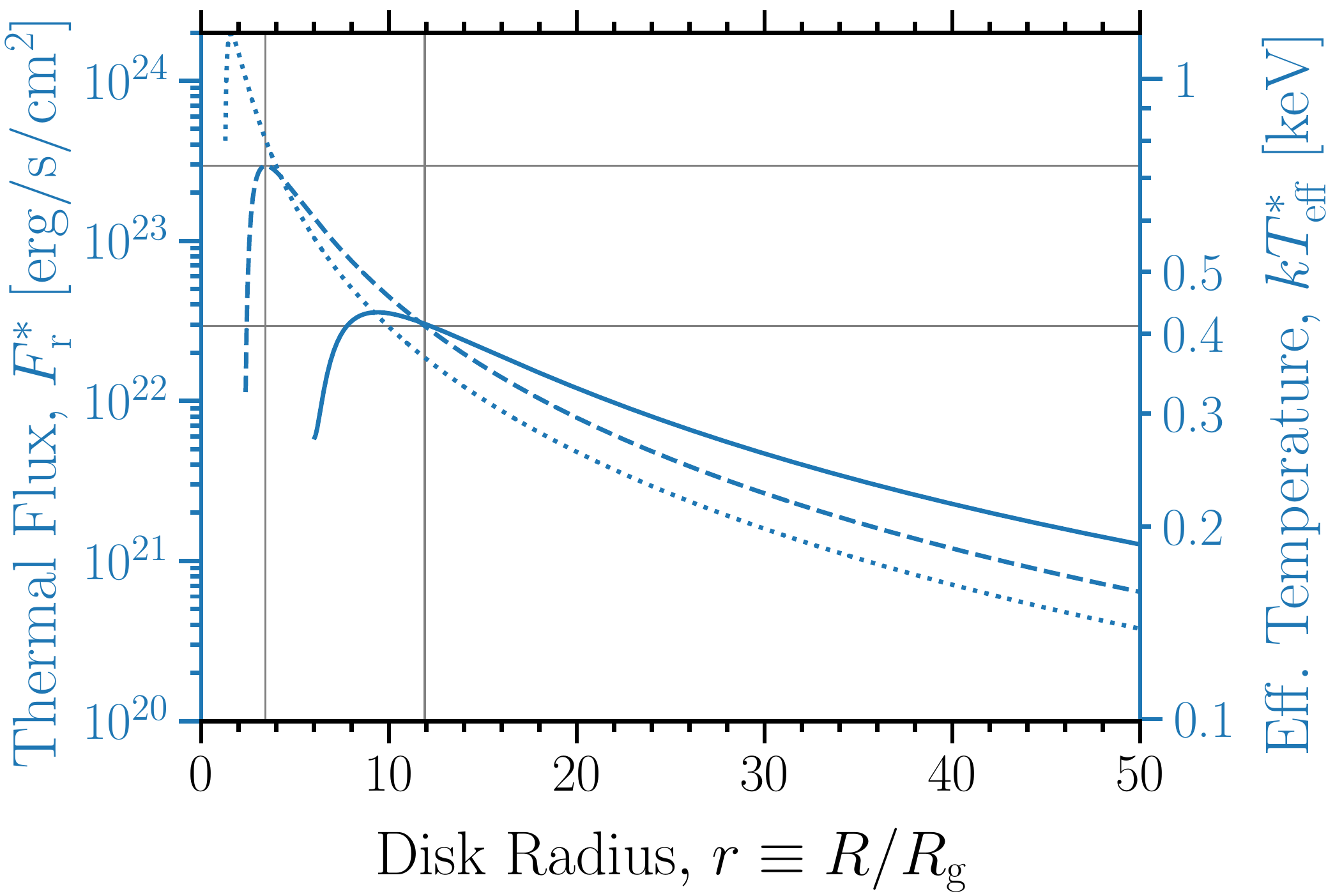}
    \vspace{-6mm}
    \caption{Total thermal radiative flux $F_{\rm r}^{\ast}$ at the effective photosphere (\textit{left axis}) and the corresponding effective temperature $T_{\rm eff}^{\ast}$ (\textit{right axis}), as a function of disk radius. Parameter choices are the same as in Figure \ref{fig:therm}, and for BH spin $a_{\ast} = 0$ (\textit{solid}), 0.9 (\textit{dashed}), 0.998 (\textit{dotted}). Thin horizontal lines mark the energy band of the thermal flux emitted from the narrow annulus where  $t_{\rm th} > t_{\rm lt}$ for the $a_{\ast} = 0.9$ case.}
    \vspace{-0mm}
    \label{fig:thermflux}
\end{figure}

Here, we present our main result that a non-negligible thermalization timescale is feasible in the intermediate state, at least for some disk radii, but negligible in the low/hard state. While this result is qualitatively consistent with high-frequency soft lag trends in BH XRBs, the strong parameter scalings predicted by our simplified model for $t_{\rm th}$ cannot explain soft lags in detail.

Figure \ref{fig:therm} plots the radial dependence of the thermalization time delay $t_{\rm th}$ (equation \ref{eqn:thermdelay}) in units of ms, for an XRB with BH mass $m = 10$ in a representative intermediate state ($l_{\rm d} = 0.2$) and low/hard state ($l_{\rm d} = 0.02$, \textit{inset}). The scaling with disk luminosity is extreme, whereby a factor of two increase in $l_{\rm d}$ (or $\dot{m}$) increases $t_{\rm th}$ by a factor of 140. The peak $t_{\rm th}$ increases by nearly an order of magnitude from BH spin $a_{\ast} = 0$ to 0.9, and again from $a_{\ast} = 0.9$ to 0.998. We consider $\alpha = 0.2$ to be a defensible choice for BH XRBs in outburst \citep{Tetarenko2018}, but acknowledge its uncertainty. We consider $\rho_{\ast} / \rho = 0.1$ to be reasonable based on BH XRB disk atmosphere models \citep[][Figure 9]{Davis2005}, but this is also uncertain. Increasing $\alpha$ or decreasing $\rho_{\ast} / \rho$ by a factor of two increases $t_{\rm th}$ seven-fold.

The key result of Figure \ref{fig:therm} is to demonstrate that the thermalization timescale plausibly rivals the light-travel timescale in the intermediate state, but is inconsequential in the low/hard state. However, our crude model predicts that $t_{\rm th} > t_{\rm lt}$ only within a narrow annulus of inner disk radii ($3.4 < r < 12$ for the $a_{\ast} = 0.9$ example).

For the same representative intermediate state parameters, Figure \ref{fig:thermflux} shows the effective temperature $T_{\rm eff}^{\ast}$ (\textit{right axis}) corresponding to the total thermal radiative flux $F_{\rm r}^{\ast}$ at the effective photosphere (\textit{left axis}; equation \ref{eqn:Teff_ast}), as a function of disk radius. Again using $a_{\ast} = 0.9$ as an example, the disk annulus with $t_{\rm th} > t_{\rm lt}$ produces thermal emission with an effective temperature range 0.41--0.73 keV. For a typical $f_{\rm col} = 1.7$, this corresponds to color temperatures in the energy band 0.70--1.2 keV, whereas the anomalously long-duration soft lags can extend down to even softer energies ($\sim 0.3~{\rm keV}$). Therefore, although our crude model suggests thermalization timescales become important in the intermediate state, our predicted $t_{\rm th}$ cannot explain soft lags in detail.

Comparing Figures \ref{fig:therm} and \ref{fig:thermflux} shows that the peak radial location of the thermalization time delay is near that of the accretion power dissipation profile. But because $t_{\rm th} \propto r^{-53/6}$ falls off so much more steeply than $F_{\rm r}^{\ast} \propto r^{-3}$, the flux-weighted thermalization time delay quickly becomes irrelevant beyond the peak radius. Effectively, this means that the thermalization reverberation lag, which is the \textit{observable} manifestation of $t_{\rm th}$, is only relevant over a narrow energy band near where the disk emission peaks. To explain the anomalously long-duration soft lags associated with larger disk radii, the thermalization timescale would need to be comparable to or exceed the light-travel timescale at these radii.

\begin{figure}[!t]
    \centering
    \vspace{-0mm}
    \includegraphics[width=0.495\textwidth]{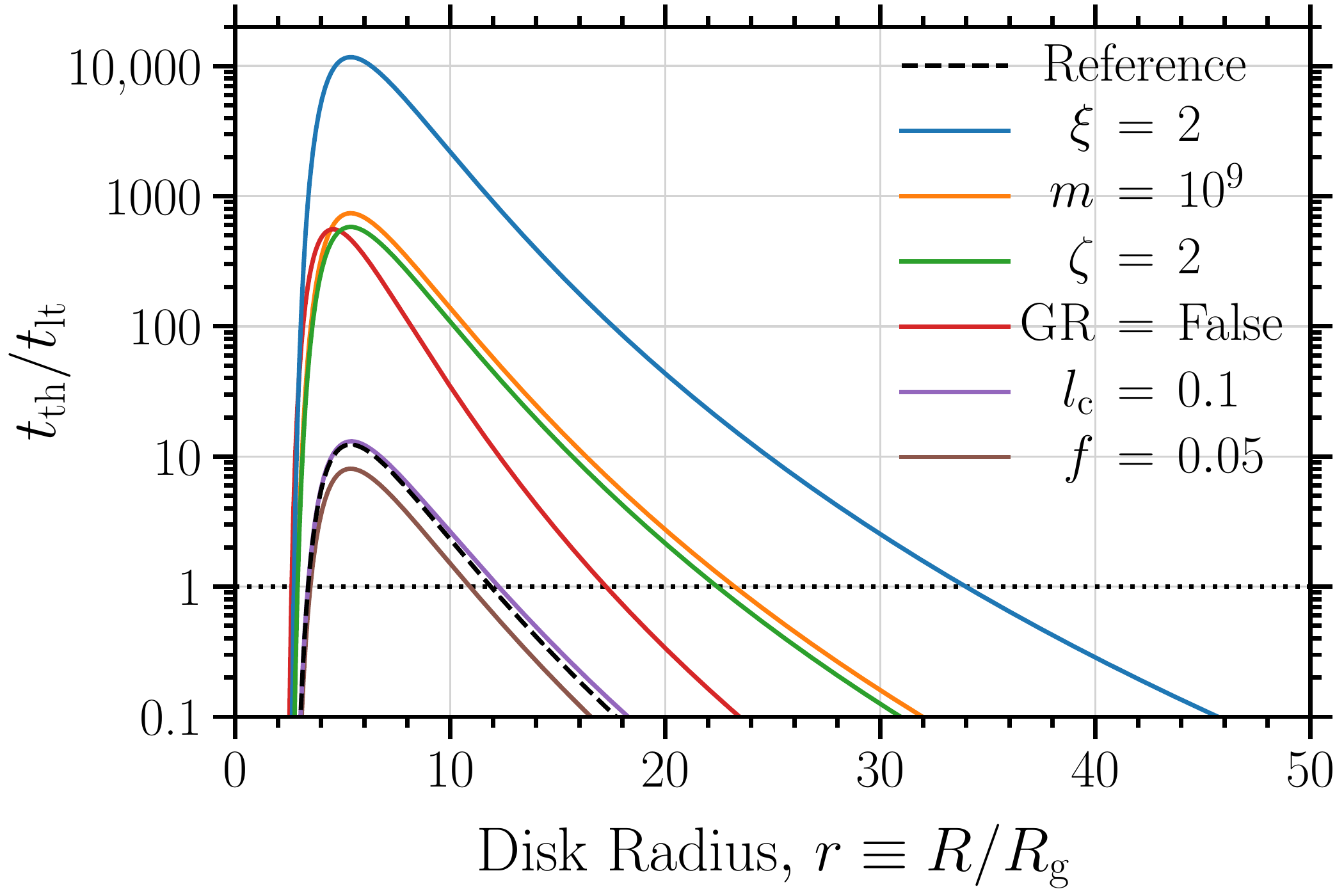}
    \vspace{-6mm}
    \caption{Thermalization time delay $t_{\rm th}$ relative to the light-travel time delay $t_{\rm lt}$ ($i = 0^{\circ}$), as a function of disk radius. Changing one parameter in the $a_{\ast} = 0.9$ reference model of Figure \ref{fig:therm} (\textit{black dashed}) isolates its effect on $t_{\rm th} / t_{\rm lt}$: no relativistic corrections (\textit{red}), $\zeta = 2$ (\textit{green}), $\xi = 2$ (\textit{blue}), $f = 0.05$ (\textit{brown}), $l_{\rm c} = 0.1$ (\textit{purple}), and $m = 10^{9}$ (\textit{orange}).}
    \vspace{-0mm}
    \label{fig:param}
\end{figure}

Figure \ref{fig:param} shows the effects of varying a single parameter in our $a_{\ast} = 0.9$ reference model. Notably, significant differences in innermost disk structure between the $\alpha$-disk model with and without relativistic corrections affect the peak $t_{\rm th} / t_{\rm lt}$ by a factor of 45. Conventions for the vertically-averaged equations for hydrostatic equilibrium \eqref{eqn:hydrostateq} and radiative diffusion \eqref{eqn:raddiffavg} strongly affect $t_{\rm th}$, with $\zeta = 2$ and $\xi = 2$ increasing $t_{\rm th}$ by factors of  47 and 950, respectively. In this sense, perhaps by adopting a relativistic disk structure and fiducial parameters $\zeta = \xi = 1$, we are conservatively estimating $t_{\rm th}$.

Radiation magnetohydrodynamics simulations of thin disks supported by radiation pressure measure accretion power dissipation in the optically thin surface layers at the few per cent level \citep{Jiang2014}, shooting up to ${\gtrsim}50\%$ for magnetic pressure-dominated support \citep{Jiang2019}. Our simple model suggests that even a small amount of dissipation in the disk surface layers ($f = 0.05$) significantly decreases $t_{\rm th}$ (factor of $1.5$), and substantial dissipation ($f = 0.5$) renders $t_{\rm th}$ irrelevant. Doubling the coronal luminosity $l_{\rm c}$ hardly affects $t_{\rm th}$.

\section{Discussion}
\label{sec:disc}
The relevance of an electron-scattering time delay to X-ray reverberation lags is unclear. During a typical hard-to-intermediate state transition, the Eddington-scaled disk luminosity $l_{\rm d}$ evolves from being undetected to $l_{\rm d} \sim 0.1$ \citep{Dunn2010, YanYu2015}. Encouragingly, our simplified model suggests that the thermalization timescale $t_{\rm th}$ can evolve from being negligible in the low/hard state, to rivaling or exceeding the light-travel timescale $t_{\rm lt}$ in the intermediate state of BH XRBs. Discouragingly, Kramers opacities and $\alpha$-disk scalings conspire to predict a steep radial decline for $t_{\rm th}$, whereas the light-travel time gradually increases with disk radius. Consequently, our predicted thermalization contribution to the reverberation lag would be isolated to a narrow inner disk annulus, corresponding to a narrow energy range near the peak disk emission (${\sim}1~{\rm keV}$). However, the energy range of interest for addressing the anomalously long soft lags is ${\sim}0.3$--$1~{\rm keV}$, mapping to larger disk radii where we predict $t_{\rm th}$ to be negligible.

Despite its structural uncertainty, our model suggests a plausibly significant thermalization timescale. The earliest attempt to account for a thermalization time delay found $t_{\rm th} \sim 70 \pm 30~{\rm ms}$ associated with a 2.2 Hz quasi-periodic oscillation from GRS 1915+105 \citep{Nathan2022}. To better understand how/if an electron-scattering time delay affects X-ray reverberation lags, we need higher-fidelity models of irradiated BH accretion disk atmospheres that include accurate opacities.

The \textit{Neutron Star Interior Composition Explorer} (\textit{NICER}) X-ray telescope collects spectral data with 100-ns timing precision \citep{Gendreau2016}, and revolutionized XRB reverberation studies \citep{Kara2019}. \textit{NICER} observed MAXI J1820+070 (hereafter J1820) throughout its outburst in 2018. As J1820 got brighter, the high-frequency soft lag associated with thermal disk reprocessing increased in duration, too long to be explained by a light-travel time delay alone \citep{Wang2021}. A systematic analysis of ${\sim}10$ BH (and candidate) XRBs confirms this as a generic trend of the hard-to-intermediate state transition, interpreted as coronal expansion/ejection \citep{Wang2022}. A thermalization time delay contribution to the soft reverberation lag would lessen the role of coronal expansion/ejection.

Independent of the soft lag evolution, evidence for a dynamic coronal geometry comes from subtle changes in the iron K emission line profile at different epochs in the rising hard state of J1820 \citep{Buisson2019}. Presumably, changes in coronal height drive changes in the disk emissivity profile, and therefore in the iron line profile. These changes are seen in spectra averaged over the duration of an observation (${\sim}10~{\rm ks}$), so should not depend on disk reprocessing timescales. Thus, an electron-scattering time delay does not imply a static corona, but may still be important for interpreting the soft lag.

GX 339--4 lends credence to an increasing electron-scattering time delay during the hard-to-intermediate state transition, with evidence for a ${\sim}100$-fold increase in the electron number density $n_{e}$ in the high/soft state compared to the low/hard state \citep{JJiang2019}. Increasing $n_{e}$ leads to more scattering events in the thermalization process; thus, a longer thermalization delay.

Interestingly, the high-frequency soft lag initially gets shorter in the rising hard state \citep[e.g.,][]{DeMarco2015}\footnote{Similarly, at the end of an outburst, the high-frequency soft lag gets longer in the decaying hard state \citep{DeMarco2017}.}, before getting longer during the intermediate state transition and evolving from ${\lesssim}1~{\rm ms}$ to ${\sim}10~{\rm ms}$ \citep{Wang2021, Wang2022}. Some groups interpret the initially decreasing soft lag as coronal contraction \citep{Kara2019, Wang2022}, whereas others argue for a decreasing inner disk truncation radius \citep{DeMarco2015, DeMarco2021, Mahmoud2019}. Because the thermalization delay is probably negligible for low mass accretion rates, we cannot settle this hard state debate.

In addition to a \textit{timing} delay effect, an electron scattering-dominated disk atmosphere causes \textit{spectral} hardening of the disk continuum, an effect approximated by a color correction factor $f_{\rm col}$ \citep[see \S\ref{sec:f_col};][]{ShimuraTakahara1995b, Davis2005}. Accounting for spectral hardening, which applies to BH accretion disks in both XRBs and active galactic nuclei (AGNs), reduces the severity of both disk truncation \citep{Salvesen2013, ReynoldsMiller2013} and the AGN disk size `problem' inferred from reverberation lags \citep{Hall2018}.

The AGN disk size problem is based on ultraviolet/optical disk continuum reverberation mapping of relatively large disk radii \citep[$r \sim 10^{2}$--$10^{4}$;][]{Cackett2021}. While our model predicts the thermalization time delay is important in the inner regions of AGN disks (see Figure \ref{fig:param}, \textit{orange line}), its applicability to larger disk radii is unclear for the reasons we discussed above.

\section{Summary and Conclusions}
\label{sec:sumconc}
BH XRBs exhibit high-frequency soft reverberation lags, associated with the thermal component of the reflection spectrum, whose durations increase from ${\lesssim}1~{\rm ms}$ to ${\sim}10~{\rm ms}$ during the hard-to-intermediate state transition \citep{Wang2022}. To explain these long-duration soft lags, we hypothesize an electron-scattering time delay for the accretion disk atmosphere to reprocess the coronal irradiation. We restrict our focus to the thermalization time delay $t_{\rm th}$ associated with the thermal response relevant to the soft lag. Based on rough but reasonable approximations, we model $t_{\rm th}$ as the cumulative time delay from individual scattering events in the random walk during the thermalization process, assuming Kramers opacities and an $\alpha$-disk model. For typical BH XRB parameters, we predict $t_{\rm th}$ is negligible in the hard state, but in the intermediate state can achieve ${\sim}10~{\rm ms}$ durations that exceed the ${\sim}1~{\rm ms}$ light-travel time delay (see Figure \ref{fig:therm}); thus, potentially lessening the role of a dynamic coronal geometry interpretation.

Ultimately, the flux-weighted contribution of the electron-scattering time delay at each disk radius determines its relevance to X-ray reverberation lags. Our crude model predicts that the thermalization contribution to the total reverberation lag can be significant at energies where the disk continuum peaks (${\sim}1~{\rm keV}$), but not at the soft lag energies of interest (${\sim}0.3~{\rm keV}$). To address the uncertain radial dependence of disk reprocessing timescales, we encourage time-dependent models of irradiated disk atmospheres with accurate opacities.

Finally, we speculate that if X-ray reverberation lag models incorporate a low-energy turnover to the irradiating spectrum, this might also contribute to increasing the soft lag duration, as follows. The soft lag comes from thermally reprocessed emission at relatively large disk radii where the light-travel time delay is long, perhaps even ${\sim}10~{\rm ms}$. But the total observed specific flux $S\(E, t\) = F\(E, t\) + R\(E, t\)$ gets swamped by the unphysical low-energy divergence of the irradiating spectrum $F\(E,t\)$, causing the small-amplitude thermal flux response $R\(E,t\)$ to be inconsequential. The argument of the cross-spectrum $G\(E, \nu\) = S\(E, \nu\) F_{\rm ref}^{\ast}\(\nu\)$ then tends to identify modulations from the corona, rather than from the disk-reprocessed thermal emission. Thus, by removing the low-energy divergence of the irradiating source spectrum, soft lags might naturally appear with long durations from light-travel time delays alone.

\section*{Acknowledgements}
GS extends his appreciation to the anonymous reviewer for an inspiring and constructive report. GS credits the idea for this paper to conversations with J. Drew Hogg in 2017 during the Kavli Institute for Theoretical Physics program, \textit{Confronting MHD Theories of Accretion Disks with Observations}, at the University of California, Santa Barbara. GS thanks Jonah M. Miller for discussing $\alpha$-disk relativistic corrections. GS thanks Bryan Kaiser for conversations on model consistency.

Research presented in this paper received support from the Laboratory Directed Research and Development program of Los Alamos National Laboratory (LANL) under project number 20220087DR. LANL approved this for unlimited release (LA-UR-22-25906).

\textit{Software}: \texttt{Python} \citep{python3}, \texttt{NumPy} \citep{numpy}, \texttt{Matplotlib} \citep{matplotlib}, \texttt{Astropy} \citep{Astropy2013, Astropy2018, Astropy2022}.

\textit{Data Availability}: \texttt{Python} scripts that generate Figures \ref{fig:therm}--\ref{fig:param} are on \texttt{GitHub}.\footnote{\href{https://github.com/gregsalvesen/thermdelay}{https://github.com/gregsalvesen/thermdelay}} LANL approved this code for open source distribution under the BSD-3 License.

\vspace{0mm}
\bibliographystyle{aasjournal}
\bibliography{salvesen}

\label{lastpage}
\end{document}